\documentclass{aa}  

\usepackage{graphicx,textcomp}
\usepackage[varg]{txfonts}

\usepackage{hyperref}
\hypersetup{colorlinks=true, urlcolor=blue, citecolor=blue}

\usepackage{color}
\usepackage{ulem}

\begin{document}

\title{Spectral properties of bright deposits in permanently shadowed craters on Ceres}
\titlerunning{Spectral properties of shadowed bright deposits on Ceres}

\author{Stefan~Schröder \inst{1}\fnmsep\thanks{Corresponding author, \email{stefanus.schroder@ltu.se}} \and Norbert~Schörghofer \inst{2} \and Erwan Mazarico \inst{3} \and Uri~Carsenty \inst{4}}

\institute{Luleå University of Technology (LTU), 98128 Kiruna, Sweden \and Planetary Science Institute (PSI), Tucson AZ 85719, USA \and NASA Goddard Space Flight Center, Greenbelt MD 20771, USA \and German Aerospace Center (DLR), 12489 Berlin, Germany}

\abstract
{Bright deposits in permanently shadowed craters on Ceres are thought to harbor water ice. However, the evidence for water ice presented thus far is indirect.}
{We aim to directly detect the spectral characteristics of water ice in bright deposits present in permanently shadowed regions (PSRs) in polar craters on Ceres.}
{We analyzed narrowband images of four of the largest shadowed bright deposits acquired by the Dawn Framing Camera  to reconstruct their reflectance spectra, carefully considering issues such as in-field stray light correction and image compression artifacts.}
{The sunlit portion of a polar deposit known to harbor water ice has a negative (blue) spectral slope of $-58 \pm 12$~\%~\textmu m$^{-1}$ relative to the background in the visible wavelength range. We find that the PSR bright deposits have similarly blue spectral slopes, consistent with a water ice composition. Based on the brightness and spectral properties, we argue that the ice is likely present as particles of high purity. Other components such as phyllosilicates may be mixed in with the ice. Salts are an unlikely brightening agent given their association with cryovolcanic processes, of which we find no trace.}
{Our spectral analysis strengthens the case for the presence of water ice in permanently shadowed craters on Ceres.}


\maketitle


\section{Introduction} \label{sec:intro}

Permanently shadowed regions (PSRs) on the Moon and Mercury are thought to harbor volatiles such as water ice \citep{A79,S92,F01,L17}. PSRs have also been identified in the polar regions of the dwarf planet Ceres \citep{S16}. \citet{P16} found bright deposits in a handful of craters identified as PSRs that they argued harbor water ice. Subsequently, \citet{E17} demonstrated that only the most persistent PSRs contain bright deposits, which strongly suggests that bright deposits in PSRs harbor volatiles accumulated in cold traps. \citet{P16} inferred that bright deposits in PSRs harbor water ice based on the spectral detection of water ice in the sunlit portion of a bright deposit in a single suspected PSR. But the fact that part of the bright deposit was seen basking in sunlight implies that the presence of water ice is unrelated to any permanent shadows on the crater floor. Using more accurate shape models, \citet{S24} showed that the extent of several bright deposits closely aligns with PSRs for different values of the tilt of the rotational axis, strongly supportive of the idea that the PSRs act as cold traps. Furthermore, the authors found that water ice is the most likely volatile to act as a  brightening agent, as the PSRs are too warm to trap supervolatiles.

While by definition direct sunlight does not enter a PSR, the inside of a PSR crater on the dayside of Ceres can be illuminated indirectly by light coming from sunlit areas on the crater wall or rim. A sufficiently sensitive camera can therefore distinguish details on the shadowed crater floor. In this study we attempted to directly detect the presence of water ice in Ceres PSRs by means of spectral characterization of the bright deposits as seen by the Dawn Framing Camera (FC). Water ice was directly detected in lunar PSRs with the help of the diagnostic absorption bands in the near-IR ($>$1~\textmu m; \citealt{L18}), although no distinct albedo markings that can be attributed to the presence of volatiles have been observed to date \citep{R24}. The FC, however, was sensitive in seven narrow bands over the 0.4-1.0~\textmu m wavelength range \citep{S11}. In this range, the spectrum of water ice generally has a negative (blue) spectral slope, with steeper slopes for larger particle sizes \citep{C81,R16,S21}. The FC was sufficiently sensitive to resolve details on shadowed crater floors, even when using the narrowband (color) filters. We analyzed FC narrowband images of bright deposits in PSRs to reconstruct their visible spectrum, looking for the tell-tale sign of a ``blue'' (negative) spectral slope. Analyzing the low reflectance values in shadowed craters requires special care because FC narrowband images were degraded by substantial in-field stray light that is difficult to remove \citep{S11,S14}. In addition, the images used in our analysis were compressed with a lossy algorithm prior to transmission, which can lead to visible artifacts depending on the compression ratio and the image content.

\begin{figure}
        \centering
        \resizebox{\hsize}{!}{\includegraphics{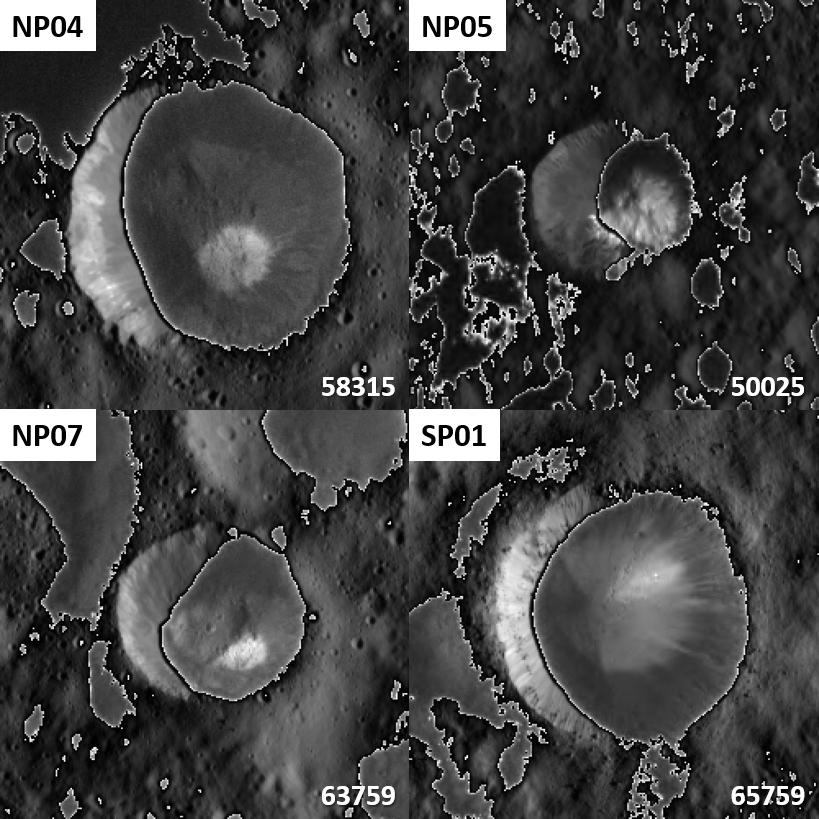}}
        \caption{Bright deposits in PSRs shown at the highest available resolution in FC clear filter images, with the image number indicated. Images are shown as unprojected $250 \times 250$ pixel-sized crops from the Low Altitude Mapping Orbit (LAMO; \citealt{R07}) with the contrast enhanced in shadowed areas.}
        \label{fig:BD_clear}
\end{figure}

\section{Method}

We determined the spectral properties of bright deposits on the shadowed floor of craters in map-projected, narrowband images. We did this by comparing the average reflectance of the deposits with that of background terrain, the latter defined as terrain that is equally shadowed but not bright. Typically, we chose the background terrain to be next to the bright deposit in the same crater or in a nearby one. Our task is hindered by the fact that the reflectance in shadowed terrain is extremely low, deriving from indirect illumination by sunlit crater walls. Apart from a low signal-to-noise, this has two other consequences. The first is that compression artifacts may dominate the signal. The narrowband images that we analyzed were compressed with the lossy 32-bit Tap algorithm \citep{S11}. Enhancing the image contrast to look into the shadow of craters can reveal compression artifacts. Whether such artifacts are noticeable depends on the image complexity and the applied compression ratio. We find that, typically, artifacts are obvious when the compression ratio is larger than 3. The second consequence of a very low reflectance is that residual stray light may contribute significantly to the reflectance. FC narrowband images are affected by in-field stray light, which is additive and varies over the field of view \citep{S11}. Filters F4, F6, F7, and F8 are affected substantially worse than filters F2, F3, and F5 (Table~\ref{tab:FC_filters}). We applied the stray light correction to the narrowband images that was developed by \citet{S14}. But this correction is first-order and leaves residual stray light that can add significantly to the very low reflectance found in shadowed terrain. Thus, when interpreting the results of the spectral analysis, we must pay close attention to compression ratio and possible stray-light artifacts.

We analyzed the spectral properties of the four largest and brightest deposits listed in \citet{E17}: NP04, NP05, NP07, and SP01 (Fig.~\ref{fig:BD_clear}). Two additional bright deposits (NP19 and NP26) are too faint to be included in our analysis. Other, smaller features identified as bright deposits by \citet{P16} are discussed in the Appendix. We calculated the reflectance of deposit and background as the average of the (projected) pixels inside selected square areas. The reflective properties of the material in a bright deposit can be reconstructed as the ratio or difference of the spectra of deposit and background. We plot the reflectance as a function of the effective wavelength ($\lambda_{\rm eff}$) of the FC narrowband filters (Table~\ref{tab:FC_filters}). If $\mu_{\rm D}(\lambda_{\rm eff}) \pm \sigma_{\rm D}(\lambda_{\rm eff})$ and $\mu_{\rm B}(\lambda_{\rm eff}) \pm \sigma_{\rm B}(\lambda_{\rm eff})$ are the average reflectance ($\pm$ standard deviation) of the pixels in the square areas over deposit and background, respectively, then the ratio spectrum is defined as $\mu_{\rm D} / \mu_{\rm B}$ with standard deviation $\sigma_{\rm D} / \mu_{\rm B}$. For reference, we also plot the ratio spectrum $\mu_{\rm B} / \mu_{\rm B} = 1$ with standard deviation $\sigma_{\rm B} / \mu_{\rm B}$. The advantage of the ratio spectrum is that it minimizes the consequences of the reflective properties of the crater wall, which, in the absence of direct sunlight, is the source of light that illuminates the deposits. Ordinarily, the ratio spectrum would provide a good estimate for the properties of the bright material, but the difference spectrum may be more representative in the case of significant stray-light artifacts. This follows from the assumption that terrain in close vicinity will suffer from a very similar degree of residual stray-light. The difference spectrum is defined as $\mu_{\rm D} - \mu_{\rm B}$ with standard deviation $(\sigma_{\rm D}^2 + \sigma_{\rm B}^2)^{1/2}$.

\begin{table}
        \centering
        \caption{Characteristics of the FC narrowband filters.}
        \label{tab:FC_filters}
        \begin{tabular}{lllll}
                \hline
                \hline
                Filter & $\lambda_{\rm cen}$ & $\Delta\lambda$ & $\lambda_{\rm eff}$ & $f$ \\
                & (nm) & (nm) & (nm) & \\
                \hline
                F2 & 549 & 44  & $555^{+15}_{-28}$ & 0.06 \\
                F3 & 749 & 45  & $749^{+22}_{-22}$ & 0.05 \\
                F4 & 919 & 46  & $917^{+24}_{-21}$ & 0.10 \\
                F5 & 978 & 87  & $965^{+56}_{-29}$ & 0.05 \\
                F6 & 829 & 37  & $829^{+18}_{-18}$ & 0.12 \\
                F7 & 650 & 43  & $653^{+18}_{-24}$ & 0.10 \\
                F8 & 428 & 41  & $438^{+10}_{-30}$ & 0.10 \\
                \hline
        \end{tabular}
        \tablefoot{$\lambda_{\rm cen}$ and $\lambda_{\rm eff}$ are the filter band center and effective wavelength, respectively, and $\Delta\lambda$ is the FWHM of the transmission profile, the boundaries of which are those indicated for $\lambda_{\rm eff}$ \citep{S13}. $f$ is the fractional contribution of in-field stray light to the average image reflectance \citep{S14}.}
\end{table}

\section{Data}

We calibrated FC2 narrowband images to reflectance ($I/F$; radiance factor) according to \citet{S13}, and subtracted in-field stray light according to \citet{S14}. We projected the calibrated images to the equirectangular map projection at a resolution of 65~pixels per degree of latitude with the USGS Integrated Software for Imagers and Spectrometers ISIS3 \citep{A04,B12}. Projection accuracy was improved by prior image registration, which involved shifting the observed image to match a simulated image prior to projection through cross-correlation and cubic interpolation, where the simulated image was created from a shape model \citep{Pr16} and the Akimov disk function as photometric model \citep{S17}. The full dataset of narrowband images used in our analysis is given in Table~\ref{tab:data_set}. The images for deposits NP04, NP05, and NP07 were acquired in the High Altitude Mapping Orbit (HAMO; \citealt{R07}), whereas those for deposit SP01 were acquired in the first Juling orbit (CXJ).

\begin{table}
        \centering
        \caption{Details of the FC2 narrowband images used for the spectral analysis, including compression ratios.}
        \label{tab:data_set}
        \begin{tabular}{llll}
                \hline
                \hline
                Deposit & Image \# & Filter & Ratio \\
                \hline
                NP04 & 45550 & F2 & 2.20 \\
        (Bilwis) & 45551 & F3 & 2.20 \\
                     & 45552 & F4 & 4.60 \\
                     & 45553 & F5 & 2.39 \\
                     & 45554 & F6 & 4.60 \\
                     & 45555 & F7 & 4.60 \\
                     & 45556 & F8 & 4.60 \\
                \hline
                NP05 & 44952 & F8 & 4.60 \\
         (Zatik) & 44953 & F7 & 4.60 \\
                     & 44954 & F6 & 4.60 \\
                     & 44955 & F5 & 2.23 \\
                     & 44956 & F4 & 4.60 \\
                     & 44957 & F3 & 2.20 \\
                     & 44958 & F2 & 2.20 \\
                \hline
                NP07 & 43732 & F8 & 4.60 \\
                     & 43733 & F7 & 4.60 \\
                     & 43734 & F6 & 4.60 \\
                     & 43735 & F5 & 2.29 \\
                     & 43736 & F4 & 4.60 \\
                     & 43737 & F3 & 2.20 \\
                     & 43738 & F2 & 2.20 \\
                \hline
                SP01 & 85738 & F2 & 2.20 \\
                     & 85739 & F3 & 2.20 \\
                     & 85740 & F4 & 4.60 \\
                     & 85741 & F5 & 4.60 \\
                     & 85742 & F6 & 4.60 \\
                     & 85743 & F7 & 2.20 \\
                     & 85744 & F8 & 4.60 \\
                \hline
        \end{tabular}
\end{table}

\section{Results}

Here we present the results of the spectral analysis for each of the four indirectly illuminated bright deposits (NP04, NP05, NP07, and SP01).

\begin{figure*}
        \centering
        \includegraphics[height=5cm,angle=0]{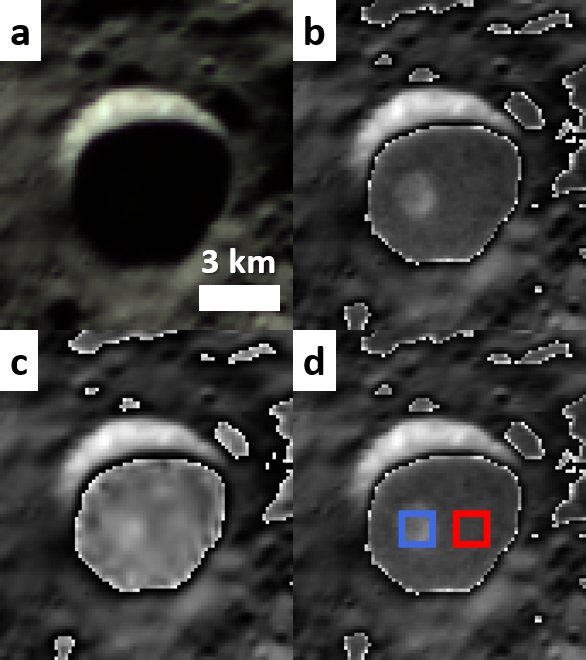}
        \includegraphics[height=5cm,angle=0]{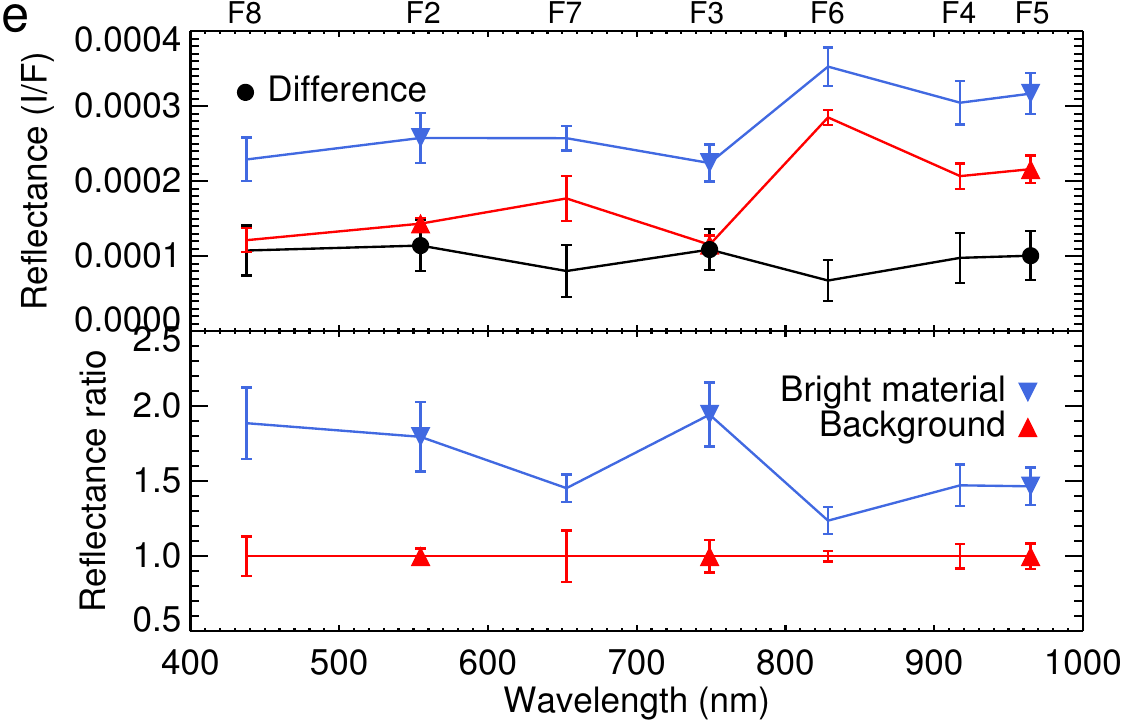}
        \caption{Bright deposit NP04 in the Bilwis crater. ({\bf a})~Color composite with (R, G, B) = (F5, F2, F8). ({\bf b})~F2 image (low compression ratio) with the contrast enhanced in shadowed areas, revealing the bright deposit. ({\bf c})~F6 image (high compression ratio) with the contrast enhanced in shadowed areas, revealing compression artifacts. ({\bf d})~F2 image in (b) with the sampled areas outlined. Outline colors correspond to the curves in (e). ({\bf e})~Spectra associated with the areas indicated in (d). Data points with a symbol are derived from images with a low compression ratio and are therefore more reliable. Top: Bright material spectra, background spectra, and their difference. Bottom: Bright material and background spectra divided by the background spectrum. The projections in (a)-(d) show the area bounded by latitudes $(85.5^\circ, 87.0^\circ)$ and longitudes $(70.0^\circ, 90.0^\circ)$E.}
        \label{fig:NP04}
\end{figure*}

{\it NP04 (Bilwis crater).} The reflectance of this bright deposit is so low that artifacts associated with high compression ratios are expected to be significant (compare Figs.~\ref{fig:NP04}b and c). Also, residual stray light may contribute significantly to the reflectance, in which case the difference between the spectra of the deposit and background may be more representative for the bright material spectrum than the ratio of the spectrum of the deposit and that of the background. We chose an area on top of the bright deposit and an equally sized background area immediately adjacent to it (Fig.~\ref{fig:NP04}d). The difference spectrum (Fig.~\ref{fig:NP04}e, top) has a neutral spectral slope for the reliable bands (low compression ratio). If we take the less reliable bands (high compression ratio) into consideration, the spectral slope is still neutral. The ratio spectrum (Fig.~\ref{fig:NP04}e, bottom) for the reliable bands seems to have a spectral slope that is not significantly different from neutral. If we include the less reliable bands, the spectral slope appears to be negative, although the scatter in the data is considerable. In short, the spectral slope of the bright deposit appears to be neutral, with weak evidence for a negative (blue) slope.

\begin{figure*}
        \centering
        \includegraphics[height=5cm,angle=0]{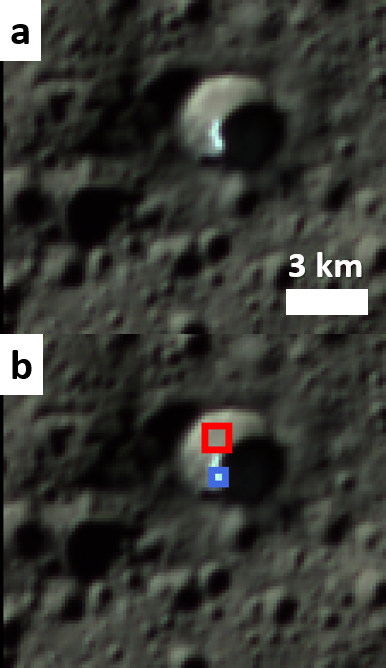}
        \includegraphics[height=5cm,angle=0]{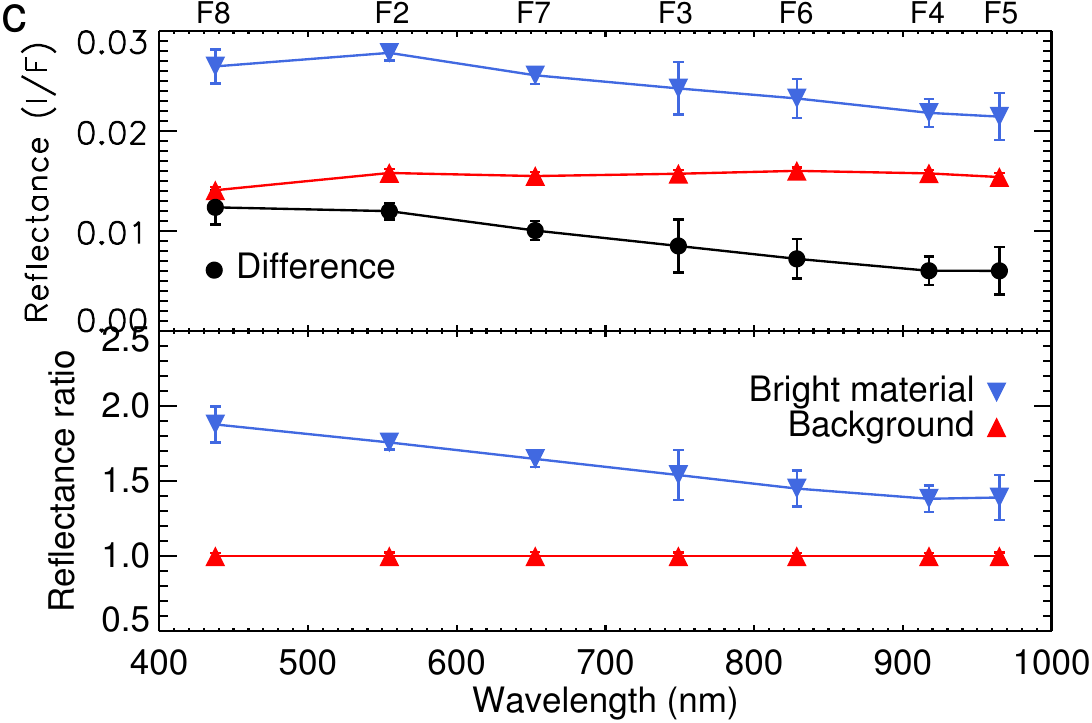}
        \caption{Sunlit bright deposit NP05 in the Zatik crater. ({\bf a})~Color composite with (R, G, B) = (F5, F2, F8). ({\bf b})~Color composite in (a) with the sampled areas outlined. Outline colors correspond to the curves in (c). ({\bf c})~Spectra associated with the areas indicated in (b). Top: Bright material spectra, background spectra, and their difference. Bottom: Bright material and background spectra divided by the background spectrum. The projections in (a) and (b) show the area bounded by latitudes $(69.0^\circ, 70.5^\circ)$ and longitudes $(111.0^\circ, 116.0^\circ)$E.}
        \label{fig:NP05_illum}
\end{figure*}

{\it NP05 (Zatik crater).} This bright deposit is located inside a crater that was recently identified to harbor a PSR \citep{S24}. The deposit is partially illuminated by the Sun, which implies that its presence is unrelated to the crater floor being permanently shadowed. The sunlit part shows spectral evidence for water ice \citep{C19}, which a recent mass wasting event may have exposed \citep{S24}. This deposit can serve as a positive control, allowing us to directly compare the spectrum of the sunlit part of the deposit with that of the shadowed deposit. The sunlit spectrum is shown in Fig.~\ref{fig:NP05_illum}. As the reflectance is so high, we need be concerned with neither the compression ratio nor the imperfect stray light correction, and it is the ratio spectrum that is relevant for the spectral shape. We chose a small area (4~projected pixels) on top of the bright deposit and a little larger background area on the crater wall adjacent to it (Fig.~\ref{fig:NP04}b). The ratio spectrum (Fig.~\ref{fig:NP05_illum}c, bottom) is consistently negative over the entire visible range with a slope\footnote{As determined with the MPFITEXY routine at \url{https://github.com/mikepqr/mpfitexy}, which depends on the MPFIT package \citep{M09}.} of $-58 \pm 12$~\% \textmu m$^{-1}$. This is the spectrum that we expect for shadowed icy deposits. Blue materials are characterized by a negative spectral slope, and, indeed, the sunlit deposit is perceived as blue in the color image (Fig.~\ref{fig:NP05_illum}a).

\begin{figure*}
        \centering
        \includegraphics[height=5cm,angle=0]{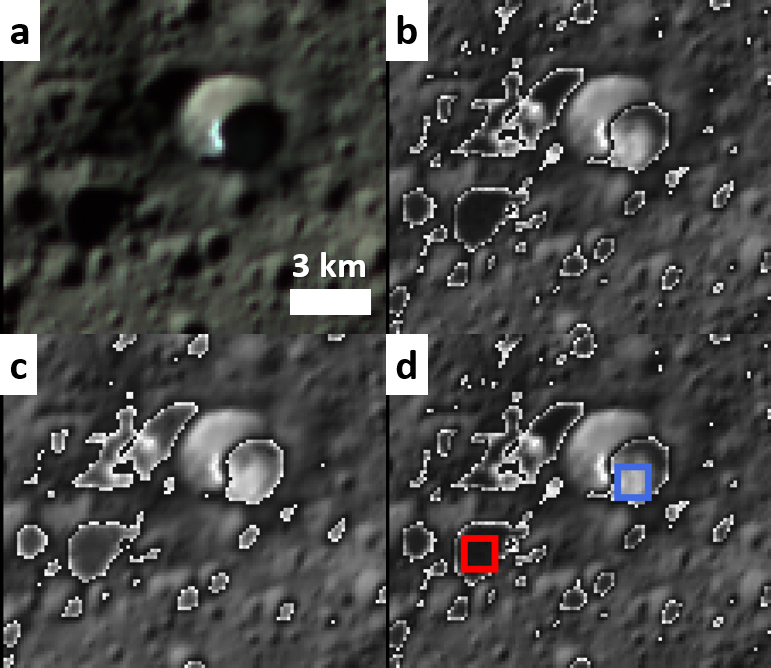}
        \includegraphics[height=5cm,angle=0]{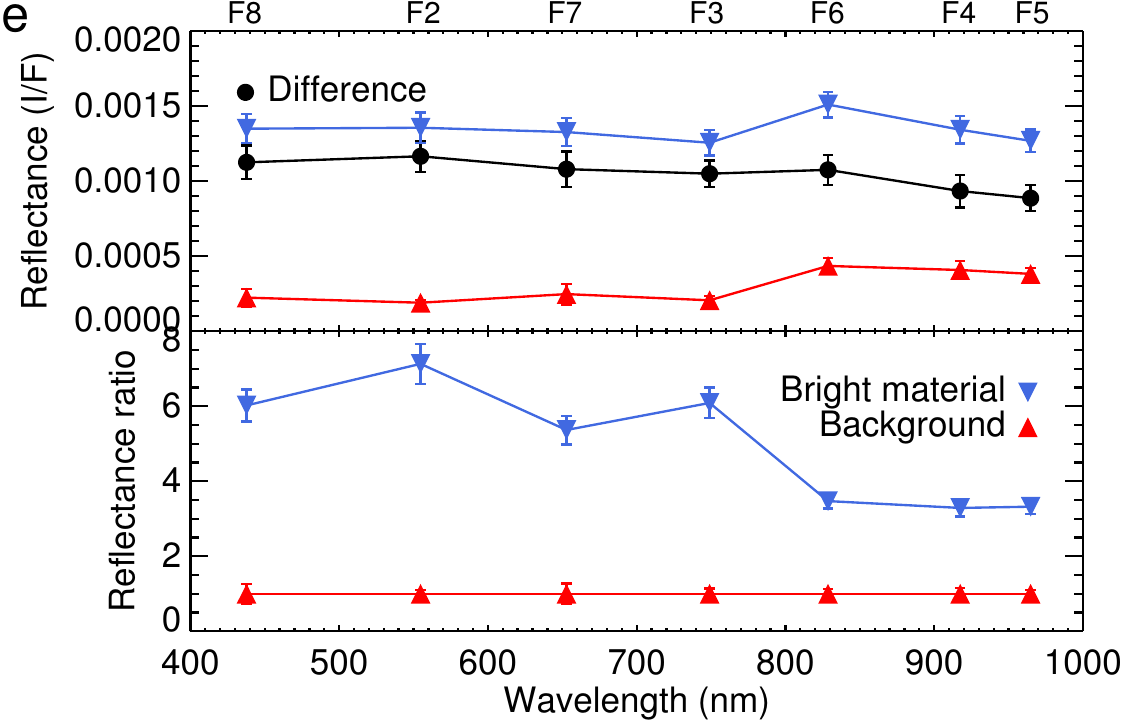}
        \caption{Shadowed bright deposit NP05 in the Zatik crater. ({\bf a})~Color composite with (R, G, B) = (F5, F2, F8). ({\bf b})~F2 image (low compression ratio) with the contrast enhanced in shadowed areas, revealing the bright deposit. ({\bf c})~F6 image (high compression ratio) with the contrast enhanced in shadowed areas. ({\bf d})~F2 image in (b) with the sampled areas outlined. Outline colors correspond to the curves in (e). ({\bf e})~Spectra associated with the areas indicated in (d). Top: Bright material spectra, background spectra, and and their difference. Bottom: Bright material and background spectra divided by the background spectrum. The projections in (a)-(d) show the area bounded by latitudes $(69.0^\circ, 70.5^\circ)$ and longitudes $(111.0^\circ, 116.0^\circ)$E.}
        \label{fig:NP05}
\end{figure*}

The spectra of the shadowed bright deposit are shown in Fig.~\ref{fig:NP05}. The reflectance of the deposit is relatively high, compared to other shadowed bright deposits, and therefore artifacts associated with high compression ratios should be minor (compare Figs.~\ref{fig:NP05}b and c). Also, residual stray light is expected to be of minor consequence. In this case, we expect the ratio of the spectrum of the deposit and that of the background to be more representative for the bright material spectrum than the difference between the spectra of the deposit and background. We chose an area on top of the bright deposit and an equally sized background area inside an adjacent crater (Fig.~\ref{fig:NP04}d). The ratio spectrum (Fig.~\ref{fig:NP04}e, bottom) has a negative (blue) spectral slope, with scatter present in the data in accordance with the size of the error bars. The slope is $-99 \pm 27$~\% \textmu m$^{-1}$, which is almost double that of the sunlit deposit, yet not significantly different given the larger uncertainty. This comparison reveals variations in slope that we can expect to find in our analysis, as inherent in the method. In any case, we conclude that we can ascertain negative spectral slopes for shadowed icy deposits.

\begin{figure*}
        \centering
        \includegraphics[height=5cm,angle=0]{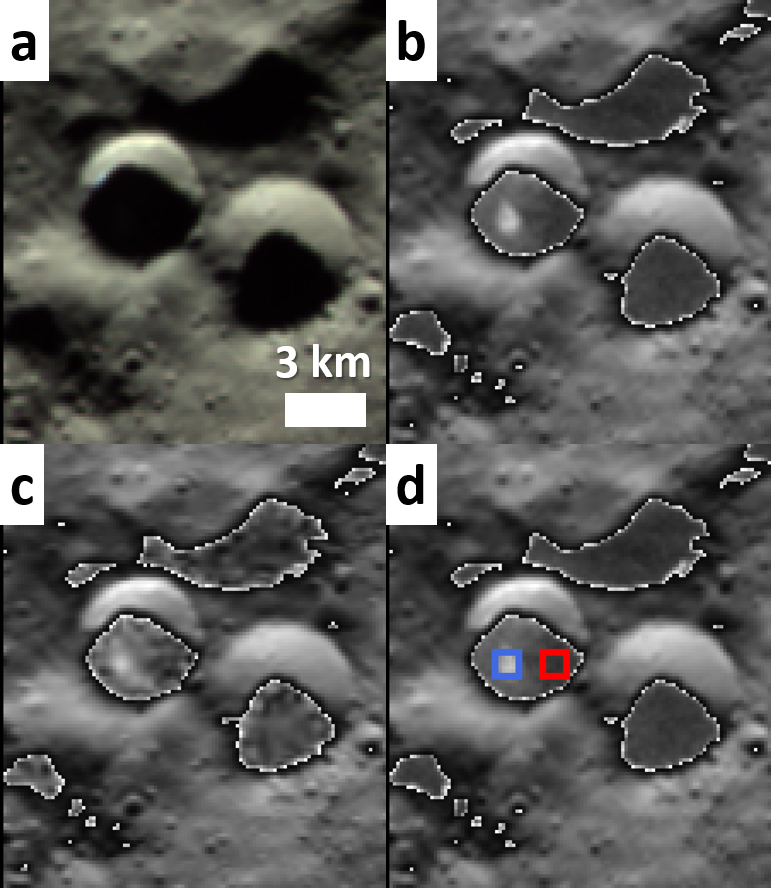}
        \includegraphics[height=5cm,angle=0]{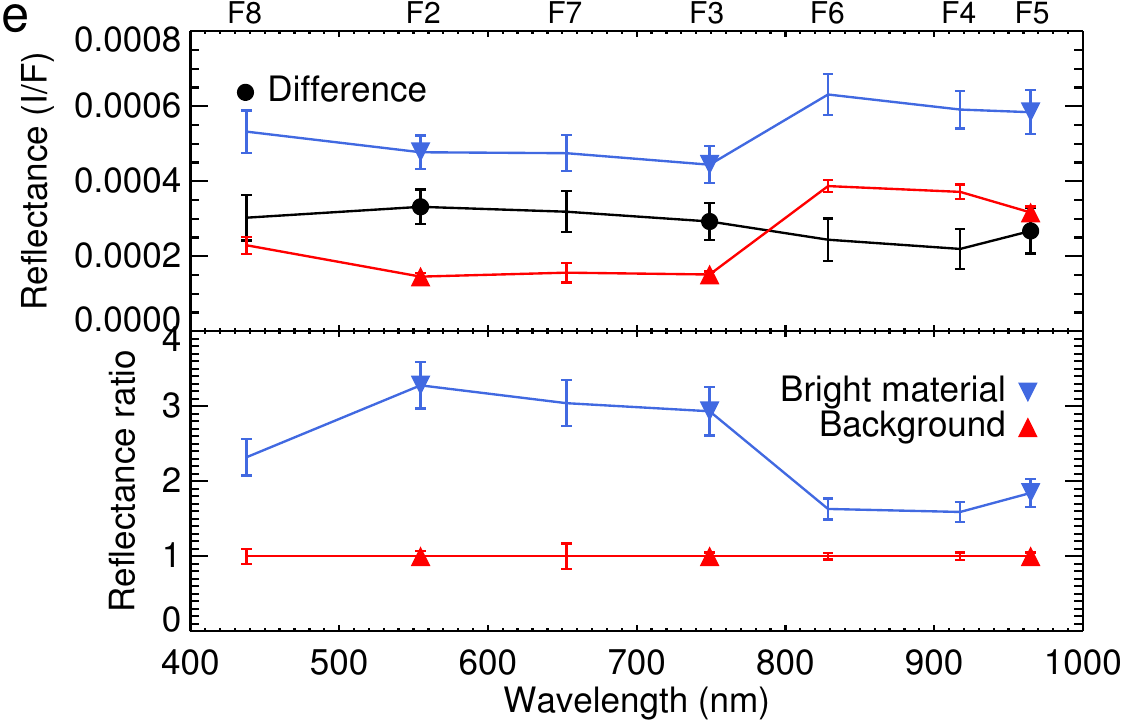}
        \caption{Bright deposit NP07. ({\bf a})~Color composite with (R, G, B) = (F5, F2, F8). ({\bf b})~F2 image (low compression ratio) with the contrast enhanced in shadowed areas, revealing the bright deposit. ({\bf c})~F7 image (high compression ratio) with the contrast enhanced in shadowed areas, revealing compression artifacts. ({\bf d})~F2 image in (b) with the sampled areas outlined. Outline colors correspond to the curves in (e). ({\bf e})~Spectra associated with the areas indicated in (d). Data points with a symbol are derived from images with a low compression ratio and are therefore more reliable. Top: Bright material spectra, background spectra, and and their difference Bottom: Bright material and background spectra divided by the background spectrum. The projections in (a)-(d) show the area bounded by latitudes $(76.5^\circ, 78.5^\circ)$ and longitudes $(351.0^\circ, 359.0^\circ)$E.}
        \label{fig:NP07}
\end{figure*}

{\it NP07.} The reflectance of the bright deposit is so low that artifacts associated with high compression ratios are expected to be significant (compare Figs.~\ref{fig:NP07}b and c). Residual stray light may contribute significantly to the reflectance, so the difference spectrum may be more representative than the ratio spectrum. We chose an area on top of the bright deposit and an equally sized background area adjacent to it (Fig.~\ref{fig:NP07}d). The difference spectrum (Fig.~\ref{fig:NP07}e, top) has an essentially neutral spectral slope for the reliable bands (low compression ratio). If we include the less reliable bands (high compression ratio), a weak negative spectral slope is apparent. The ratio spectrum (Fig.~\ref{fig:NP07}e, bottom) for the reliable bands appears to have a negative slope in the sense that the ratio for F5 (965~nm) is significantly lower than that of F2 (555~nm) and F3 (749~nm). Including the less reliable bands confirms the negative spectral slope, although the scatter in the data is considerable. In short, the spectral slope of the bright deposit appears to be negative (blue).

\begin{figure*}
        \centering
        \includegraphics[height=5cm,angle=0]{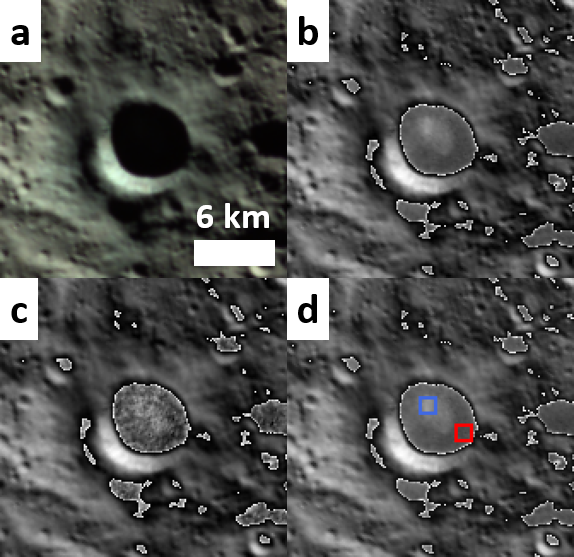}
        \includegraphics[height=5cm,angle=0]{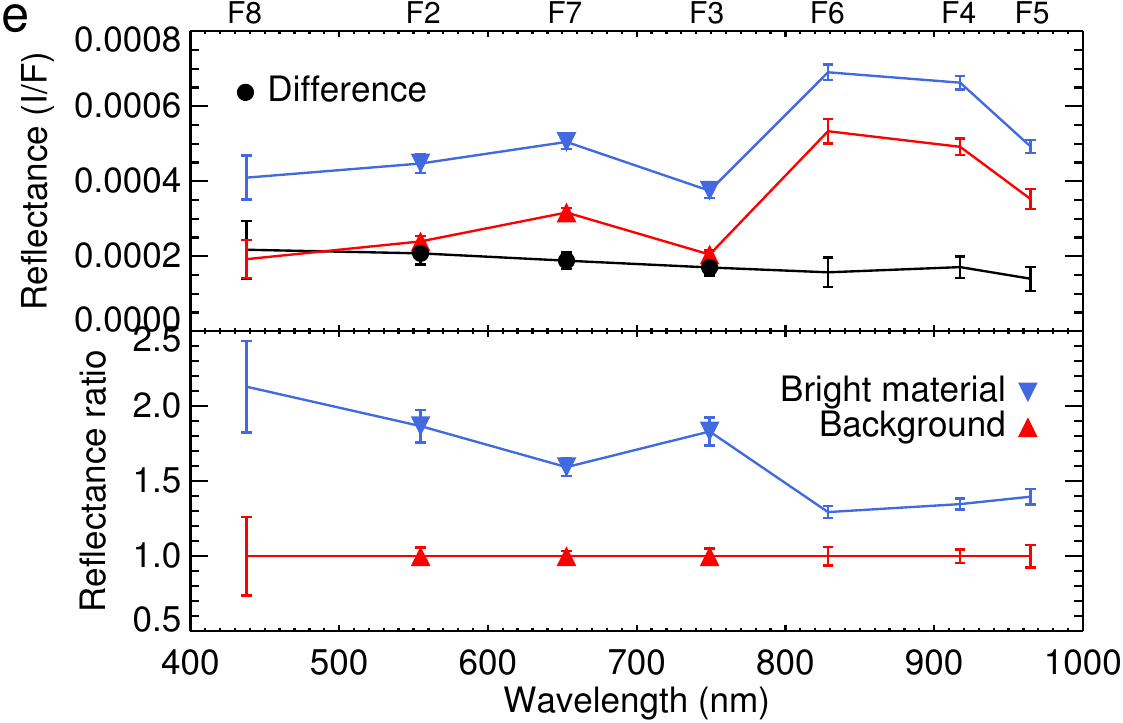}
        \caption{Bright deposit SP01. ({\bf a})~Color composite with (R, G, B) = (F5, F2, F8). ({\bf b})~F2 image (low compression ratio) with the contrast enhanced in shadowed areas, revealing the bright deposit. ({\bf c})~F8 image (high compression ratio) with the contrast enhanced in shadowed areas, revealing compression artifacts. ({\bf d})~F2 image in (b) with the sampled areas outlined. Outline colors correspond to the curves in (e). ({\bf e})~Spectra associated with the areas indicated in (d). Data points with a symbol are derived from images with a low compression ratio and are therefore more reliable. Top: Bright material spectra, background spectra, and and their difference. Bottom: Bright material and background spectra divided by the background spectrum. The projections in (a)-(d) show the area bounded by latitudes $(-72.5^\circ, -70.0^\circ)$ and longitudes $(27.0^\circ, 35.0^\circ)$E.}
        \label{fig:SP01}
\end{figure*}

{\it SP01.} The reflectance of the bright deposit is so low that artifacts associated with high compression ratios are expected to be significant (compare Figs.~\ref{fig:SP01}b and c). Residual stray light may contribute significantly to the reflectance, so the difference spectrum may be most representative. We chose an area on top of the brightest part of the bright deposit and an equally sized background area immediately adjacent to the deposit (Fig.~\ref{fig:SP01}d). The difference spectrum (Fig.~\ref{fig:SP01}e, top) has a shallow but regular negative spectral slope for the reliable bands (low compression ratio). Including the less reliable bands (high compression ratio) confirms the negative spectral slope. The ratio spectrum (Fig.~\ref{fig:SP01}e, bottom) for the reliable bands has a neutral slope. If we include the less reliable bands, the spectral slope appears to be negative, although the scatter in the data is considerable. In short, the spectral slope of the bright deposit is likely negative (blue).

\section{Discussion}

\begin{figure}
        \centering
        \resizebox{\hsize}{!}{\includegraphics{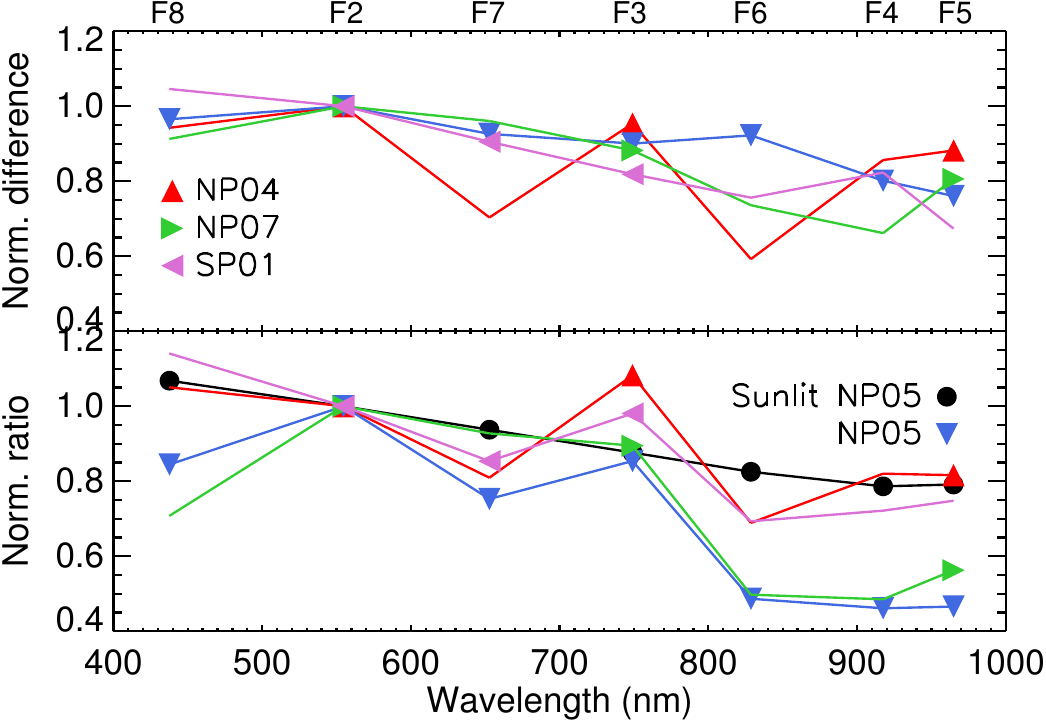}}
        \caption{Normalized reflectance ratio and normalized reflectance difference for the four bright deposits. Spectra are normalized at 549~nm (filter F2). Error bars have been omitted for clarity. Data points with a symbol are derived from images with a low compression ratio and are therefore more reliable.}
        \label{fig:all_combi}
\end{figure}

While some bright deposits clearly show the spectral signature for water ice in the form of a negative spectral slope, for others the evidence is only marginal. Figure~\ref{fig:all_combi} summarizes the results from Figs.~\ref{fig:NP05_illum} to \ref{fig:SP01} by showing the (normalized) ratio and difference of the spectra of bright and background terrain for all four deposits. Ordinarily, the ratio would be the appropriate indicator for the spectral properties of the deposits, but in this case -- with a very low reflectance in combination with possible stray light artifacts -- the difference may be more representative. The sunlit deposit NP05 indicates that surface ice is expected to feature a regular negative spectral slope. A negative trend can be seen in both the ratio and difference spectra of PSR deposits NP04, NP07, and SP01, albeit with considerable scatter in the data. Due to the low signal-to-noise in the shadowed terrain, we cannot establish with statistical certainty that each individual deposit has a negative spectral slope, but the fact that all three deposits have negative trends leads us to conclude that, in general, bright deposits in PSRs exhibit a negative spectral slope in the visible. Several spectra in Fig.~\ref{fig:all_combi} also appear to feature absorption bands around 650 and 830~nm. However, the reality of these bands cannot be established with certainty given that they are associated with filters F7 and F6, which are among the most severely affected by in-field stray light.

On Ceres, ``blue'' material featuring a negative spectral slope is commonly found in young terrain, but it is not bright \citep{SK16,S17}. On the other hand, typical bright terrains on Ceres are salt deposits, which feature a spectral slope ranging from neutral to positive \citep{dS16,S17,dS24}. Additionally, salts on Ceres are believed to derive from brines from the interior that extruded onto the surface in a process known as cryovolcanism. Salts have been reported in the interior of very large craters that show evidence of cryovolcanism, such as Occator and Dantu \citep{dS16,dS24}. Salt deposits have not been reported for small craters like our PSR craters. While morphological markers for cryovolcanism can be hard to distinguish in permanently shadowed areas, we know that the impacts that create small craters do not deposit enough energy to result in brine extrusion from the interior \citep{R20}. It is therefore highly unlikely that salts would end up in the interior of PSR craters given that the craters themselves are not obviously located on the floor of much larger ones. Therefore, the bright deposits in the PSRs are unlikely to be salts.

The fact that the deposits are relatively bright indicates that the ice is present in particulate form, as this would lead to an increased fraction of multiply scattered light and brightening \citep{AF67}. The ice particles themselves are likely relatively pure, given that only minor amounts of a darkening agent in an intimate mixture can lead to a large drop in reflectance and the vanishing of spectral features \citep{AF67,CL84}. But other mineral components may be present in an areal mixture with the ice, with phyllosilicates an obvious candidate as they are ubiquitous on the surface \citep{A16}. Exactly to what extent other components are mixed in can be addressed with dedicated experiments or spectral modeling. The spectrum of the sunlit portion of deposit NP05, with its high signal-to-noise, would lend itself best to such a modeling effort, especially when also considering the spectrum acquired by the Dawn spectrometer \citep{C19}.

\section*{Data availability}

Dawn FC images of Ceres that are calibrated to intensity are available at the Small Bodies Node of NASA's Planetary Data System\footnote{\url{https://pds-smallbodies.astro.umd.edu/}.}. Narrowband image data for the four PSR craters in this paper, calibrated to reflectance and corrected for stray light, are available at \url{https://doi.org/10.5281/zenodo.11908582}.

\begin{acknowledgements}

The authors thank an anonymous reviewer for comments that led to improvements in the manuscript.
\end{acknowledgements}

\bibliography{Ceres}

\begin{thebibliography}{30}
\expandafter\ifx\csname natexlab\endcsname\relax\def\natexlab#1{#1}\fi

\bibitem[{{Adams} \& {Filice}(1967)}]{AF67}
{Adams}, J.~B. \& {Filice}, A.~L. 1967, \jgr, 72, 5705

\bibitem[{{Ammannito} {et~al.}(2016){Ammannito}, {DeSanctis}, {Ciarniello},
  {Frigeri}, {Carrozzo}, {Combe}, {Ehlmann}, {Marchi}, {McSween}, {Raponi},
  {Toplis}, {Tosi}, {Castillo-Rogez}, {Capaccioni}, {Capria}, {Fonte},
  {Giardino}, {Jaumann}, {Longobardo}, {Joy}, {Magni}, {McCord}, {McFadden},
  {Palomba}, {Pieters}, {Polanskey}, {Rayman}, {Raymond}, {Schenk}, {Zambon},
  \& {Russell}}]{A16}
{Ammannito}, E., {DeSanctis}, M.~C., {Ciarniello}, M., {et~al.} 2016, Science,
  353, aaf4279

\bibitem[{{Anderson} {et~al.}(2004){Anderson}, {Sides}, {Soltesz}, {Sucharski},
  \& {Becker}}]{A04}
{Anderson}, J.~A., {Sides}, S.~C., {Soltesz}, D.~L., {Sucharski}, T.~L., \&
  {Becker}, K.~J. 2004, in Lunar and Planetary Institute Science Conference
  Abstracts, Vol.~35, Lunar and Planetary Institute Science Conference
  Abstracts, ed. S.~{Mackwell} \& E.~{Stansbery}, 2039

\bibitem[{{Arnold}(1979)}]{A79}
{Arnold}, J.~R. 1979, \jgr, 84, 5659

\bibitem[{{Becker} {et~al.}(2012){Becker}, {Anderson}, {Barrett}, {Sides}, \&
  {Titus}}]{B12}
{Becker}, K.~J., {Anderson}, J.~A., {Barrett}, J.~M., {Sides}, S.~C., \&
  {Titus}, T.~N. 2012, in Lunar and Planetary Institute Science Conference
  Abstracts, Vol.~43, Lunar and Planetary Institute Science Conference
  Abstracts, 2892

\bibitem[{{Clark}(1981)}]{C81}
{Clark}, R.~N. 1981, \jgr, 86, 3087

\bibitem[{{Clark} \& {Lucey}(1984)}]{CL84}
{Clark}, R.~N. \& {Lucey}, P.~G. 1984, \jgr, 89, 6341

\bibitem[{{Combe} {et~al.}(2019){Combe}, {Raponi}, {Tosi}, {De Sanctis},
  {Carrozzo}, {Zambon}, {Ammannito}, {Hughson}, {Nathues}, {Hoffmann}, {Platz},
  {Thangjam}, {Schorghofer}, {Schr{\"o}der}, {Byrne}, {Landis}, {Ruesch},
  {McCord}, {Johnson}, {Singh}, {Raymond}, \& {Russell}}]{C19}
{Combe}, J.-P., {Raponi}, A., {Tosi}, F., {et~al.} 2019, \icarus, 318, 22

\bibitem[{{De Sanctis} {et~al.}(2024){De Sanctis}, {Ammannito}, {Carrozzo},
  {Ciarniello}, {De Angelis}, {Ferrari}, {Frigeri}, \& {Raponi}}]{dS24}
{De Sanctis}, M.~C., {Ammannito}, E., {Carrozzo}, F.~G., {et~al.} 2024,
  Communications Earth and Environment, 5, 131

\bibitem[{{De Sanctis} {et~al.}(2016){De Sanctis}, {Raponi}, {Ammannito},
  {Ciarniello}, {Toplis}, {McSween}, {Castillo-Rogez}, {Ehlmann}, {Carrozzo},
  {Marchi}, {Tosi}, {Zambon}, {Capaccioni}, {Capria}, {Fonte}, {Formisano},
  {Frigeri}, {Giardino}, {Longobardo}, {Magni}, {Palomba}, {McFadden},
  {Pieters}, {Jaumann}, {Schenk}, {Mugnuolo}, {Raymond}, \& {Russell}}]{dS16}
{De Sanctis}, M.~C., {Raponi}, A., {Ammannito}, E., {et~al.} 2016, \nat, 536,
  54

\bibitem[{{Ermakov} {et~al.}(2017){Ermakov}, {Mazarico}, {Schr{\"o}der},
  {Carsenty}, {Schorghofer}, {Preusker}, {Raymond}, {Russell}, \&
  {Zuber}}]{E17}
{Ermakov}, A.~I., {Mazarico}, E., {Schr{\"o}der}, S.~E., {et~al.} 2017, \grl,
  44, 2652

\bibitem[{{Feldman} {et~al.}(2001){Feldman}, {Maurice}, {Lawrence}, {Little},
  {Lawson}, {Gasnault}, {Wiens}, {Barraclough}, {Elphic}, {Prettyman},
  {Steinberg}, \& {Binder}}]{F01}
{Feldman}, W.~C., {Maurice}, S., {Lawrence}, D.~J., {et~al.} 2001, \jgr, 106,
  23231

\bibitem[{{Lawrence}(2017)}]{L17}
{Lawrence}, D.~J. 2017, Journal of Geophysical Research (Planets), 122, 21

\bibitem[{{Li} {et~al.}(2018){Li}, {Lucey}, {Milliken}, {Hayne}, {Fisher},
  {Williams}, {Hurley}, \& {Elphic}}]{L18}
{Li}, S., {Lucey}, P.~G., {Milliken}, R.~E., {et~al.} 2018, Proceedings of the
  National Academy of Science, 115, 8907

\bibitem[{{Markwardt}(2009)}]{M09}
{Markwardt}, C.~B. 2009, in Astronomical Society of the Pacific Conference
  Series, Vol. 411, Astronomical Data Analysis Software and Systems XVIII, ed.
  D.~A. {Bohlender}, D.~{Durand}, \& P.~{Dowler}, 251

\bibitem[{{Platz} {et~al.}(2016){Platz}, {Nathues}, {Schorghofer}, {Preusker},
  {Mazarico}, {Schr{\"o}der}, {Byrne}, {Kneissl}, {Schmedemann}, {Combe},
  {Sch{\"a}fer}, {Thangjam}, {Hoffmann}, {Gutierrez-Marques}, {Landis},
  {Dietrich}, {Ripken}, {Matz}, \& {Russell}}]{P16}
{Platz}, T., {Nathues}, A., {Schorghofer}, N., {et~al.} 2016, Nature Astronomy,
  1, 0007

\bibitem[{{Preusker} {et~al.}(2016){Preusker}, {Scholten}, {Matz}, {Elgner},
  {Jaumann}, {Roatsch}, {Joy}, {Polanskey}, {Raymond}, \& {Russell}}]{Pr16}
{Preusker}, F., {Scholten}, F., {Matz}, K.-D., {et~al.} 2016, in Lunar and
  Planetary Science Conference, Vol.~47, Lunar and Planetary Science
  Conference, 1954

\bibitem[{{Raponi} {et~al.}(2016){Raponi}, {Ciarniello}, {Capaccioni},
  {Filacchione}, {Tosi}, {De Sanctis}, {Capria}, {Barucci}, {Longobardo},
  {Palomba}, {Kappel}, {Arnold}, {Mottola}, {Rousseau}, {Quirico}, {Rinaldi},
  {Erard}, {Bockelee-Morvan}, \& {Leyrat}}]{R16}
{Raponi}, A., {Ciarniello}, M., {Capaccioni}, F., {et~al.} 2016, \mnras, 462,
  S476

\bibitem[{{Raymond} {et~al.}(2020){Raymond}, {Ermakov}, {Castillo-Rogez},
  {Marchi}, {Johnson}, {Hesse}, {Scully}, {Buczkowski}, {Sizemore}, {Schenk},
  {Nathues}, {Park}, {Prettyman}, {Quick}, {Keane}, {Rayman}, \&
  {Russell}}]{R20}
{Raymond}, C.~A., {Ermakov}, A.~I., {Castillo-Rogez}, J.~C., {et~al.} 2020,
  Nature Astronomy, 4, 741

\bibitem[{{Robinson} {et~al.}(2024){Robinson}, {Mahanti}, {Bussey}, {Carter},
  {Denevi}, {Estes}, {Humm}, {Kincyzk}, {Li}, {Lucey}, {Mazarico}, {Ravine},
  {Speyerer}, {Wagner}, \& {Williams}}]{R24}
{Robinson}, M.~S., {Mahanti}, P., {Bussey}, D.~B.~J., {et~al.} 2024, in LPI
  Contributions, Vol. 3040, LPI Contributions, 1669

\bibitem[{{Russell} {et~al.}(2007){Russell}, {Capaccioni}, {Coradini}, {de
  Sanctis}, {Feldman}, {Jaumann}, {Keller}, {McCord}, {McFadden}, {Mottola},
  {Pieters}, {Prettyman}, {Raymond}, {Sykes}, {Smith}, \& {Zuber}}]{R07}
{Russell}, C.~T., {Capaccioni}, F., {Coradini}, A., {et~al.} 2007, Earth Moon
  and Planets, 101, 65

\bibitem[{{Schmedemann} {et~al.}(2016){Schmedemann}, {Kneissl}, {Neesemann},
  {Stephan}, {Jaumann}, {Krohn}, {Michael}, {Matz}, {Otto}, {Raymond}, \&
  {Russell}}]{SK16}
{Schmedemann}, N., {Kneissl}, T., {Neesemann}, A., {et~al.} 2016, \grl, 43,
  11,987

\bibitem[{{Schorghofer} {et~al.}(2024){Schorghofer}, {Gaskell}, {Mazarico}, \&
  {Weirich}}]{S24}
{Schorghofer}, N., {Gaskell}, R., {Mazarico}, E., \& {Weirich}, J. 2024, \psj,
  5, 99

\bibitem[{{Schorghofer} {et~al.}(2016){Schorghofer}, {Mazarico}, {Platz},
  {Preusker}, {Schr{\"o}der}, {Raymond}, \& {Russell}}]{S16}
{Schorghofer}, N., {Mazarico}, E., {Platz}, T., {et~al.} 2016, \grl, 43, 6783

\bibitem[{{Schr{\"o}der} {et~al.}(2013){Schr{\"o}der}, {Maue}, {Guti{\'e}rrez
  Marqu{\'e}s}, {Mottola}, {Aye}, {Sierks}, {Keller}, \& {Nathues}}]{S13}
{Schr{\"o}der}, S.~E., {Maue}, T., {Guti{\'e}rrez Marqu{\'e}s}, P., {et~al.}
  2013, \icarus, 226, 1304

\bibitem[{{Schr{\"o}der} {et~al.}(2017){Schr{\"o}der}, {Mottola}, {Carsenty},
  {Ciarniello}, {Jaumann}, {Li}, {Longobardo}, {Palmer}, {Pieters}, {Preusker},
  {Raymond}, \& {Russell}}]{S17}
{Schr{\"o}der}, S.~E., {Mottola}, S., {Carsenty}, U., {et~al.} 2017, \icarus,
  288, 201

\bibitem[{{Schr{\"o}der} {et~al.}(2014){Schr{\"o}der}, {Mottola}, {Matz}, \&
  {Roatsch}}]{S14}
{Schr{\"o}der}, S.~E., {Mottola}, S., {Matz}, K.~D., \& {Roatsch}, T. 2014,
  \icarus, 234, 99

\bibitem[{{Sierks} {et~al.}(2011){Sierks}, {Keller}, {Jaumann}, {Michalik},
  {Behnke}, {Bubenhagen}, {B{\"u}ttner}, {Carsenty}, {Christensen}, {Enge},
  {Fiethe}, {Guti{\'e}rrez Marqu{\'e}s}, {Hartwig}, {Kr{\"u}ger}, {K{\"u}hne},
  {Maue}, {Mottola}, {Nathues}, {Reiche}, {Richards}, {Roatsch},
  {Schr{\"o}der}, {Szemerey}, \& {Tschentscher}}]{S11}
{Sierks}, H., {Keller}, H.~U., {Jaumann}, R., {et~al.} 2011, \ssr, 163, 263

\bibitem[{{Slade} {et~al.}(1992){Slade}, {Butler}, \& {Muhleman}}]{S92}
{Slade}, M.~A., {Butler}, B.~J., \& {Muhleman}, D.~O. 1992, Science, 258, 635

\bibitem[{{Stephan} {et~al.}(2021){Stephan}, {Ciarniello}, {Poch}, {Schmitt},
  {Haack}, \& {Raponi}}]{S21}
{Stephan}, K., {Ciarniello}, M., {Poch}, O., {et~al.} 2021, Minerals, 11, 1328

\end{thebibliography}
\bibliographystyle{aa} 

\begin{appendix}
        
        \section{Other proposed shadowed bright deposits}
        
        In this work we analyzed the spectral properties of four bright deposits in PSRs listed in \citet{E17}. In addition, seven smaller ($< 3$~km diameter) deposits were proposed by \citet{P16}. We evaluated clear-filter images from LAMO and discuss the seven candidates in the order in which they are listed in \citet{P16} (\#4-\#10).
        
\begin{figure*}
                \centering
                \includegraphics[width=\textwidth,angle=0]{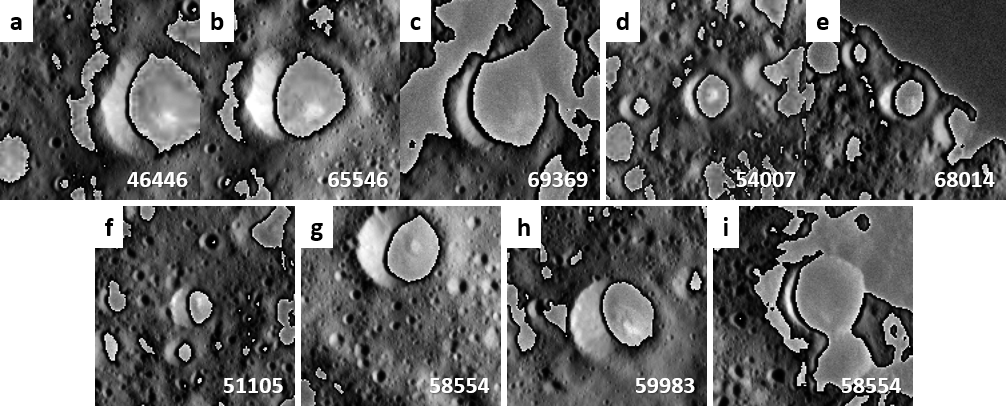}
                \caption{Additional bright deposits in shadowed craters proposed by \citet{P16} shown at the highest available resolution in clear filter LAMO images, with the FC image number indicated. Images are shown as unprojected $150 \times 150$ pixel-sized crops with the contrast enhanced in shadowed areas.  ({\bf a}), ({\bf b}), and ({\bf c}) deposit \#4. ({\bf d}) and ({\bf e}) \#5. ({\bf f}) \#7. ({\bf g}) \#8. ({\bf h}) \#9. ({\bf i}) \#10.}
                \label{fig:small_BDs}
        \end{figure*}

        {\it \#4}. This proposed bright deposit is associated with a relatively large crater (3.1~km crater diameter). It was originally identified in image 46446, which has a lossy compression ratio of 3.60. This ratio is relatively high and compression artifacts are apparent in the shadows in the form of blurred, blob-like brightness variations (Fig.~\ref{fig:small_BDs}a). We evaluated two other LAMO images that were compressed with the relatively low ratio of 2.50. There appears to be a bright deposit present in image 65546 (Fig.~\ref{fig:small_BDs}b). But even though the compression ratio is relatively low, compression artifacts are still noticeable. The same crater in image 69369, which has a lower degree of indirect illumination, does not harbor the bright feature seen in image 65546 (Fig.~\ref{fig:small_BDs}c). Visible in the shadows are subtle reflectance variations that may be bright streaks or result from topography. The conflicting observational evidence leads us to conclude that the status of bright deposit \#4 is unresolved. Modeling reveals that this crater hosts a PSR, making the presence of a bright ice deposit in this crater plausible \citep{S24}.

        {\it \#5}. This proposed bright deposit was originally identified in image 54007, which has a lossy compression ratio of 3.60. This ratio is relatively high and enhancing shadowed areas reveals compression artifacts (Fig.~\ref{fig:small_BDs}d). The same crater in image 68014, which has a compression ratio of 2.50, does not harbor the bright feature in the center of the shadowed area as seen in image 46446 (Fig.~\ref{fig:small_BDs}e). Visible in the shadows is what appears to be a bright streak on the crater wall such as those that can be seen in sunlit craters on Ceres. Bright deposit \#5 may be a compression artifact.

        {\it \#6}. This proposed bright deposit was never a serious candidate. It was identified as a very small group of bright pixels. Such groups result from topography and pop up frequently in any image sufficiently stretched in brightness to see into the shadows.
        
        {\it \#7}. This proposed bright deposit was originally identified in image 51105 (Fig.~\ref{fig:small_BDs}f). The nature of this deposit is unclear because of its small size, but may be a bright streak on the crater wall such as those that can be seen in sunlit craters on Ceres.
        
        {\it \#8}. This proposed bright deposit was originally identified in image 58554 (Fig.~\ref{fig:small_BDs}g). The nature of this deposit is unclear because of its small size, but may either result from topography or be a genuine, albeit very faint, bright deposit.
        
        {\it \#9}. This proposed bright deposit was originally identified in image 59983 (Fig.~\ref{fig:small_BDs}h). The nature of this deposit is unclear because of its small size, but may be a bright streak on the crater wall such as those that can be seen in sunlit craters on Ceres. The streak is located in an area that appears to have slumped into the crater as part of a possible landslide.
        
        {\it \#10}. This proposed bright deposit was originally identified in image 58554 (Fig.~\ref{fig:small_BDs}i). The nature of this deposit is unclear because of its small size, but may be a faint streak on the crater wall such as those that can be seen in sunlit craters on Ceres.
        
\end{appendix}



\end{document}